\documentclass[a4paper]{article}

\usepackage{INTERSPEECH2022}

\usepackage{multirow,multicol}

\usepackage{hyperref}
\usepackage{cleveref}

\usepackage{color,soul}

\newcommand{\tfidf}{TF-IDF}
\newcommand{\tfidfpun}{TF-IDF-PUN}
\newcommand{\bert}{BERT}
\newcommand{\bertpun}{BERT-PUN}
\newcommand{\roberta}{roBERTa}
\newcommand{\robertapun}{roBERTa-PUN}
\newcommand{\wordtovec}{Word2Vec}
\newcommand{\fasttext}{FastText}
\newcommand{\glove}{GloVe}

\title{Automatic Detection of Expressed Emotion from Five-Minute Speech Samples: Challenges and Opportunities}
\name{Bahman Mirheidari$^{1*}$, André Bittar$^{2*}$, Nicholas Cummins$^2$, Johnny Downs$^2$, \\Helen L. Fisher$^2$, Heidi Christensen$^1$ \thanks{Submitted to Interspeech 2022}}
\address{
  $^*$Joint First Authors \\
  $^1$Department of Computer Science, University of Sheffield, UK \\
  $^2$Institute of Psychiatry, Psychology \& Neuroscience (IoPPN), King’s College London, London, UK}
\email{b.mirheidari@sheffield.ac.uk, andre.bittar@kcl.ac.uk}

\begin{document}

\maketitle
\begin{abstract} 
 
\noindent 
We present a novel feasibility study on the automatic recognition of Expressed Emotion (EE), a family environment concept based on caregivers speaking freely about their relative/family member. We describe an automated approach for determining the \textit{degree of warmth}, a key component of EE, from acoustic and text features acquired from a sample of 37 recorded interviews. These recordings, collected over 20 years ago, are derived from a nationally representative birth cohort of 2,232 British twin children and were manually coded for EE. We outline the core steps of extracting usable information from recordings with highly variable audio quality and assess the efficacy of four machine learning approaches trained with different combinations of acoustic and text features. Despite the challenges of working with this legacy data, we demonstrated that the degree of warmth can be predicted with an $F_{1}$-score of \textbf{61.5\%}. In this paper, we summarise our learning and provide recommendations for future work using real-world speech samples.
\end{abstract}
\vspace{0.2cm}


\section{Introduction}
Speech-based emotion recognition (SER) technologies and systems have been steadily increasing in prominence in the speech processing literature over the last two decades~\cite{Schuller_2018, swain2018databases}. Typical SER approaches focus on one of two tasks: (i) the recognition of \textit{discrete} emotions, typically the six `basic' emotions identified by Ekman~\cite{ekman1992argument}; and (ii) \textit{continuous} emotion recognition along a dimensional representation, typically arousal and valence~\cite{Kuppens13-TRB}. 
SER applications can provide feedback within supportive technologies within healthcare systems~\cite{Harley17-ATO,jaimes07-MHC}. 
For example, SER influenced applications have been proposed in early-diagnosis settings. Recently, speech markers have been used to assist in the inference of \textit{Attachment Condition} in school-age children~\cite{alsofyani21_interspeech}. This approach used an SER-style approach to recognise if children were emotionally secure or insecure. Similarly, the tracking of emotional engagement can be used to assess the negative impact of dementia on communication~\cite{steinert21_interspeech}.

The work presented herein is also based around the concept of automatic speech-based emotion recognition within a clinical application. Specifically, we present the analysis of \textit{Expressed Emotion} (EE), a family environment concept based on caregivers speaking freely about the relative/family member in their care~\cite{kanter1987expressed, hibbs1991determinants}. We focus on the automatic recognition of EE ratings from caregiver speech samples talking about their 5-year-old twins. To the best of the authors' knowledge, this is the first time an automated approach for determining EE in such settings has been attempted. Adding to the challenges of this work is the quality of the audio recordings: they were originally made 20 years ago on audio cassette tapes. 
Despite the associated challenges, we are able to demonstrate that \textsl{degree of warmth}, a component of EE, can be predicted with 61.5\% $F_{1}$-score. 

The rest of the paper is organised as follows: Section~\ref{bg} presents the relevant background from the fields of psychiatry and psychology, Section~\ref{expsetup} describes the set-up, data, and methods of this study, Section~\ref{results} contains the results, and concluding remarks are provided in Section 5. 

\section{Background}\label{bg}

In the field of psychiatry, EE refers to the attitudes of caregivers towards their child and comprises criticism, hostility, and/or emotional over-involvement, as well as the degree of warmth shown. For over five decades, levels of EE within families have been studied by psychologists and psychiatrists to determine which adults with mental illness are likely to have the poorest outcomes~\cite{brown_influence_1972, rutter_reliability_1966, Vaughn1976TheMO}. EE was originally measured through in-depth face-to-face interviews but, due to time constraints, has subsequently been assessed through brief samples of caregivers speaking freely about their child. These interactions are known as the \textit{Five-Minute Speech Sample} (FMSS)~\cite{magana_brief_1986}.

Coding of EE from these easy-to-collect speech samples focuses on the emotions that are apparent when the caregiver speaks about their child, drawing on both what is said and the tone of voice. This coding can contain clinically useful information. For example, EE rated from maternal speech samples plays a causal role in the development of antisocial behavioural problems in children\cite{caspi_maternal_2004} and subsequent serious mental illnesses.\cite{moffitt_males_2002} Other studies have shown that ratings of negative emotions from parents’ speech predict the onset and course of other mental health problems in children, including anxiety, depression, and attention-deficit hyperactivity disorder\cite{sher-censor_five_2015} underlining its usefulness as an early predictor of youth mental health difficulties. 
However, this promising prediction method is rarely used; 
the coding of speech is labour-intensive and requires highly trained raters. Moreover, human rating potentially has limited reproducibility as it can be prone to drift and unconscious biases. Automating the assessment of EE could dramatically impact clinical practice, and provide clinicians with an important guide to the likelihood that a young person will develop mental health problems and permit them to effectively target preventive interventions and reduce incidence rates of mental disorders.

\section{Experimental setup}\label{expsetup}


This section describes the data used within our presented analysis. We introduce the original cohort study, describe the key steps required to produce the analyzable data and outline the methods used in our experiments.

\subsection{Cohort Study}

The Environmental Risk (E-Risk) Longitudinal Twin Study tracks the development of a nationally representative birth cohort of 2,232 British twin children born in England and Wales in 1994-1995. They have been comprehensively assessed during home visits at ages 5, 7, 10, 12 and 18 years (with 93\% retention). The Joint South London and Maudsley and Institute of Psychiatry Research Ethics Committee approved each phase of the study.
When the children were 5 years old, speech samples of approximately five minutes were audio-recorded from caregivers (almost exclusively mothers) in their homes to elicit expressed emotion about each child. Trained interviewers asked caregivers to describe each of their children (“For the next 5 minutes, I would like you to describe [child] to me; what is [child] like?”). The caregiver was encouraged to talk freely but if s/he found this difficult, a series of semi-structured probes were used (e.g., “In what ways would you like [child] to be different?”). These speech samples were coded by two trained raters according to manualised guidelines with high inter-rater reliability (r=0.84–0.90). Ratings included the degree of dissatisfaction/negativity and the degree of warmth that the caregiver expressed towards each child (0=none to 5=high).

\subsection{Data}

The interviews from the E-Risk study were recorded on cassette tapes using consumer-grade equipment available at the time. The cassettes were maintained in storage for 20 years and may have degraded during this time. The audio quality is highly variable with frequent inaudible passages and white noise. The tapes required digitisation, which was carried out by an external contractor using professional equipment.
Although the interviews follow a loose structure based on the use of standard prompt questions, they contain passages of overlapping speech. Furthermore, there is often background chatter, interruptions by young children, and other ambient noises.
To enable an analysis of linguistic content and for the training of an automatic speech recogniser (ASR), the interviews had to be transcribed. The significant cost of professional transcription and the small project budget imposed limitations on the amount of material we were able to transcribe. 
Furthermore, the low quality of the audio complicated the process, and the resulting transcripts contained numerous alignment errors, inaccurate segmentation, incorrect time-stamping, missing passages and some incorrectly rendered words.
The final sample contained 37 transcribed interviews coded by human raters for EE, a small proportion of the total.

\subsection{Methods}

Our experiment aimed at to classify interview samples for the level of warmth expressed by mothers towards their twins. We sought to assess and compare the efficacy of using different combinations of features -- acoustic-only, text-only, and both of these together.


We adopted a strategy to increase the amount of available data. Using the Audacity\footnote{https://www.audacityteam.org/} audio editor, we aligned the audio and transcripts and manually tagged all utterances to indicate the speaker, which twin was being referred to, and the content type of the utterance, following the loose interview structure. For example, the tag {\tt int-both-support} indicates the interviewer asking about the level of support the mother received during pregnancy, relating to both twins. The tag {\tt mum-t1-away} indicates a mother's description of her feelings when the elder twin (t1) is away from her. The equivalent tag for the younger twin is {\tt mum-t2-away}. Our tagging schema included 38 distinct tags (19 interviewer prompt tags and 19 mother utterance tags). This tagging enabled us to double the size of the dataset by splitting each audio sample into utterances for elder and younger twins. Thus, the dataset used for our experiments contained 74 samples. The EE ratings for each twin were used as target labels for classification. 

\begin{figure}[t]
  \vspace{-0.1cm}
    \centering
    \includegraphics[width=7.5cm,height=5.5cm]{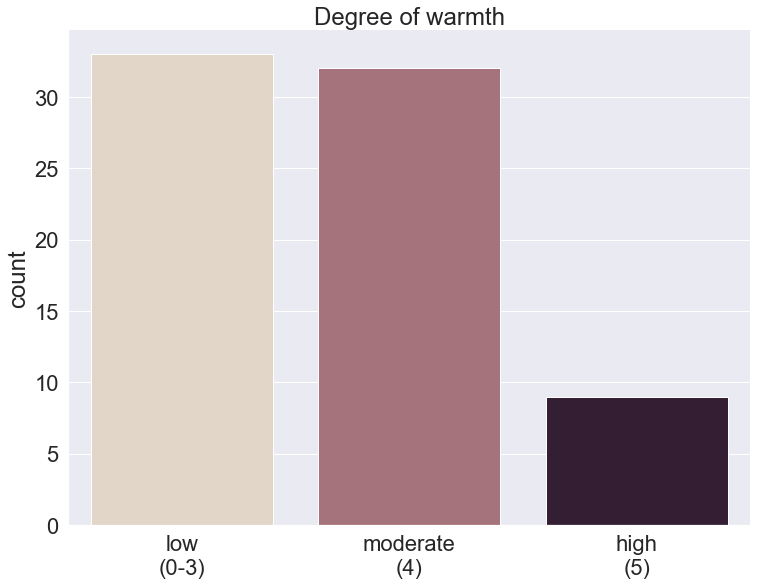}
    \caption{Distribution of caregiver warmth classes in 3-way schema with corresponding 6-way schema numerical classes.}
    \label{fig:3-way-distrib}
    \vspace{-0.2cm}
\end{figure}

The original human-based coding of warmth from the E-Risk study consisted of 6 ordinal classes ranging from 0 to 5. However, the distribution of classes was imbalanced in the available dataset. We therefore merged classes into a 3-class schema to provide a more balanced distribution of classes for training our models. The distribution of classes in the final dataset is shown in Figure~\ref{fig:3-way-distrib}. The resulting distribution, while not perfectly balanced, was more even than the original coding.


\section{Results}\label{results}

We used four different machine learning classifiers (with default parameters except for increasing the number of iterations) to predict the degree of warmth for the individual twins: \textit{Logistic Regression} (LR), \textit{Linear Support Vector Classifier} (Lin-SVC), \textit{Random Forest} (RF), and \textit{K-Nearest Neighbours} (KNN). We used the Scikit-learn \cite{scikit-learn} library in Python to train the classifiers. To evaluate the performance of the classifiers, we ran classification tasks five times using stratified 5-fold cross-validation with shuffling. The final metric was the average of the $F_{1}$-score of the classifiers over the runs.

\subsection{Acoustic features}
We extracted different frame-based acoustic features from the caregivers' audio segments using the 
\textsc{open\-SMILE} Toolkit~\cite{eyben2010opensmile}. We took the average and standard deviation of the features to build fixed-length feature sets for training. These features were introduced by the authors of \textsc{open\-SMILE} at various Interspeech challenges on the detection of emotion and paralinguistic characteristics of speech, including the \textit{Interspeech 2009 Emotion Detection Challenge} (IS09-EMO)~\cite{schuller2009interspeech}, \textit{Audio-Visual Emotion recognition Challenge 2013} (AVEC13)~\cite{valstar2013avec},
Interspeech 2010, 2013 and 2016 \textit{Computational Paralinguistics ChallengE} (ComParE) (IS10-CMP, IS13-CMP, IS16-CMP)~\cite{schuller2010interspeech,schuller2013interspeech,schuller2016interspeech}, and the \textit{extended Geneva Minimal Acoustic Parameter Set} (eGeMAPS)~\cite{eyben2015geneva}.

Table~\ref{tab:warmth-acoustic-only} shows the average $F_{1}$-score and standard deviation of the four classifiers on different acoustic features. The results varied across combinations of features and classifiers. The RF classifier using IS-10 features and the LR classifier with eGeMAPS resulted in the best average $F_{1}$-score of 59.9\%; the former having fewer errors (2.2\% vs. 4.4\%). KNN on both IS13-CMP and IS16-CMP scored lowest ($F_{1}$=38.9\%). IS13-CMP and IS16-CMP features on all classifiers compared to IS10-CMP features had statistically significantly lower $F_{1}$-scores (p-value $ < 0.05$). In addition, the results obtained by the RF classifier were significantly better than the other classifiers.

\begin{table}
\small
\caption{Average $F_{1}$-score and standard deviation (5 runs, 5-fold cross validation) of the four classifiers using different acoustic-only features.}
\label{tab:warmth-acoustic-only}
\centering{ 
\begin{tabular}{c|c |c|c|c } 
\hline
\multicolumn{1}{c|}{\textbf{Features}} &  
\multicolumn{1}{c|}{\textbf{LR}}&
\multicolumn{1}{c|}{\textbf{Lin-SVC}} &  
\multicolumn{1}{c|}{\textbf{RF}} &
\multicolumn{1}{c}{\textbf{KNN}}\\  
\hline
AVEC13&51.6(3.5)&40.2(4.8)&\textbf{58.8(2.9)}&39.1(3.0) \\  
IS09-EMO&54.7(2.3)&52.6(2.4)&\textbf{57.6(2.5)}&45.3(2.9) \\
IS10-CMP&59.2(2.1)&59.2(1.7)&\textbf{59.9(2.2)}&55.2(2.6) \\
IS13-CMP&52.3(3.6)&43.1(3.9)&\textbf{57.1(1.5)}&38.9(2.9) \\
IS16-CMP&49.5(3.4)&43.1(3.9)&\textbf{56.6(2.0)}&38.9(2.9)  \\
eGeMAPS&\textbf{59.9(4.4)}&47.7(7.4)&58.3(2.1)&42.4(2.8)  \\
\hline
-& 54.5(4.2)&47.6(7.1)&\textbf{58.0(1.2)}&43.3(6.4) \\
\hline
\multicolumn{5}{l}{\small{The best result in a row is in bold.}} 
\end{tabular}
\vspace{-0.4cm}
}
\end{table}

\subsection{Text features}

%

We tested three methods to represent the textual content of the interviews: i) Term Frequency-Inverse document frequency using the Scikit-Learn \cite{scikit-learn} \texttt{TfIdfVectorizer} with default settings fitted to the training set (\tfidf); ii) pre-trained word emeddings: GloVe \cite{pennington2014glove} 25 dimensions, trained on Twitter data (\glove), FastText \cite{joulin2016bag} 300 dimensions, trained on Wikipedia (\fasttext), Word2Vec \cite{NIPS2013_9aa42b31} 300 dimensions, trained on the Google News corpus (\wordtovec); iii) Pre-trained transformer-based language models\footnote{We used the {\tt transformers} library by HuggingFace, https://huggingface.co/}: BERT \cite{devlin2018bert} base, large, uncased model (\bert), and roBERTa \cite{liu2019roberta} large base uncased model (\roberta).

We preprocessed the transcripts using spaCy \cite{spacy2020}, tokenising the text, removing punctuation and lingering whitespace, and lowercasing all tokens. For the pre-trained language models, which are limited to sequences of 512 tokens, the text was divided into chunks of 512 tokens and passed to the models using a sliding window approach with 50\% overlap and the average and standard deviation of the last three layers was computed to create features for classification.
For \tfidf and pre-trained language models, we trialled using the original transcripts with punctuation (reported in results with the suffix \;\; \textsl{"-pun"}) and the pre-processed texts. Word embedding models used the pre-processed text only.
For the embedding models, unknown words (words that did not appear in the model) were ignored. The final embedding representation for each transcript was the mean of embeddings for all known word tokens in the transcript. In all runs, only utterances for caregivers were retained as this was the content used for the manual coding of warmth in the transcripts. We used the same classification models for the acoustic features. 


Table \ref{tab:warmth-text-only} shows the average $F_{1}$-score and standard deviation of the four classifiers on the different text features. The best average $F_{1}$-score was achieved by a Lin-SVC classifier using the roberta-pun features ($F_{1}$=53.7\%), while the lowest performance was obtained with RF and bert features ($F_{1}$=36.5\%). The bert features on all classifiers also had significantly lower $F_{1}$-scores (p-value $< 0.05$) compared to roberta-pun features. However, the difference between the classifiers' performance was not statistically significant.

\begin{table}[t]
\small
\caption{Average $F_{1}$-score and standard deviation (5 runs, 5-fold cross validation) of the four classifiers using different text-only features.  
}
\label{tab:warmth-text-only}
\centering{ 
\begin{tabular}{c|c |c|c|c } 
\hline
\multicolumn{1}{c|}{\textbf{Features}} &  
\multicolumn{1}{c|}{\textbf{LR}}&
\multicolumn{1}{c|}{\textbf{Lin-SVC}} &  
\multicolumn{1}{c|}{\textbf{RF}} &
\multicolumn{1}{c}{\textbf{KNN}}\\  
\hline
\tfidf&\textbf{43.4(3.2)}&41.4(2.7)&42.9(5.4)&43.2(4.2)\\
\tfidfpun&39.9(3.6)&42.4(3.9)&\textbf{46.6(2.9)}&37.5(2.3)\\
\bert&37.9(2.6)&37.0(2.3)&36.5(2.8)&\textbf{41.4(3.7)}\\
\bertpun&49.2(2.9)&\textbf{53.4(3.9)}&46.4(4.7)&44.0(3.3)\\
\roberta&43.6(3.5)&43.4(6.1)&\textbf{45.9(2.9)}&41.1(1.8)\\
\robertapun& 49.9(3.3)&\textbf{53.7(2.7)}&42.3(4.2)&48.1(3.7)\\
\wordtovec&44.4(3.9)&47.5(3.1)&47.3(2.7)&\textbf{47.8(4.1)}\\
\fasttext&38.7(5.1)&42.9(4.8)&\textbf{46.7(5.5)}&40.6(2.8)\\
\glove&43.4(3.9)&48.4(3.8)&53.5(3.5)&43.5(2.2)\\
\hline
-& 43.4(4.2)&\textbf{45.6(5.6)}&45.3(5.6)&43.0(3.4) \\
\hline
\multicolumn{5}{l}{\small{The best result in a row is in bold.}} 
\end{tabular}
}
\end{table}

\begin{table}[t]
\small
\caption{Average $F_{1}$-score and standard deviation (5 runs, 5-fold cross validation) of the four classifiers using IS10-CMP acoustic features combined with different text features (Comb. Feat.).  
}
\label{tab:warmth-comb-feats}
\centering{ 
\begin{tabular}{c|c |c|c|c } 
\hline
\multicolumn{1}{c|}{\textbf{Comb. Feat.}} &  
\multicolumn{1}{c|}{\textbf{LR}}&
\multicolumn{1}{c|}{\textbf{Lin-SVC}} &  
\multicolumn{1}{c|}{\textbf{RF}} &
\multicolumn{1}{c}{\textbf{KNN}}\\  
\hline
\tfidf&59.2(2.1)&\textbf{61.0(3.2)}&53.6(3.3)&54.6(2.0)\\
\tfidfpun&59.0(2.2)&\textbf{61.5(2.3)}&52.8(2.7)&54.3(2.1)\\ 
\bertpun&\textbf{59.8(2.6)}&58.4(2.8)&45.3(4.3)&55.4(3.3)\\
\roberta&59.9(2.4)&\textbf{60.4(2.6)}&46.2(2.1)&55.1(2.1)\\
\robertapun& \textbf{58.8(2.3)}&58.0(2.8)&45.9(4.5)&55.4(2.5) \\
\wordtovec&59.2(2.1)&\textbf{59.3(2.5)}&52.9(1.7)& 54.7(2.3) \\
\fasttext&59.2(2.1)&\textbf{60.7(2.4)}&52.9(4.9)&53.5(2.5) \\
\glove&59.5(2.2)&\textbf{60.8(4.1)}&57.1(1.5)&53.8(2.8)\\
\hline
-&59.3(0.4)&\textbf{60.0(1.3)}&50.8(4.4)&54.6(0.7) \\
\hline
\multicolumn{5}{l}{\small{The best result in a row is in bold.}} 
\end{tabular}
}
\end{table} 

\subsection{Combining acoustic and text features}

The manual coding of warmth (and EE in general) relies on both interview content and voice features. We therefore sought to assess the use of both modalities in the classification task, using a combination of acoustic and text features to train the models.
Since the IS10-CMP acoustic features yielded the best results on all classifiers, we combined these features with each of the text features, with the exception of \bert, due to its significantly lower performance in the text-only task. The average $F_{1}$-score of the classifiers on the combined features are shown in Table \ref{tab:warmth-comb-feats}\footnote{Note that we tried to normalise the features before the combination, however we could not see much difference in the results, so we ignored the normalisation when combining features.}. 
The overall figures across different classifiers and features showed an increase in performance compared to both acoustic-only and text-only experiments. 
The best performance was obtained by the Lin-SVC classifier using \tfidfpun features ($F_{1}$=61.5\%). The second best performance was obtained by the same classifier using \tfidf features on the pre-processed text ($F_{1}$=61\%). A t-test showed no significant differences between the $F_{1}$-scores across all features. However, Lin-SVC and LR classifiers performed significantly better than KNN and RF.

\begin{figure}[t]
 \centering
 \includegraphics[width=8cm]{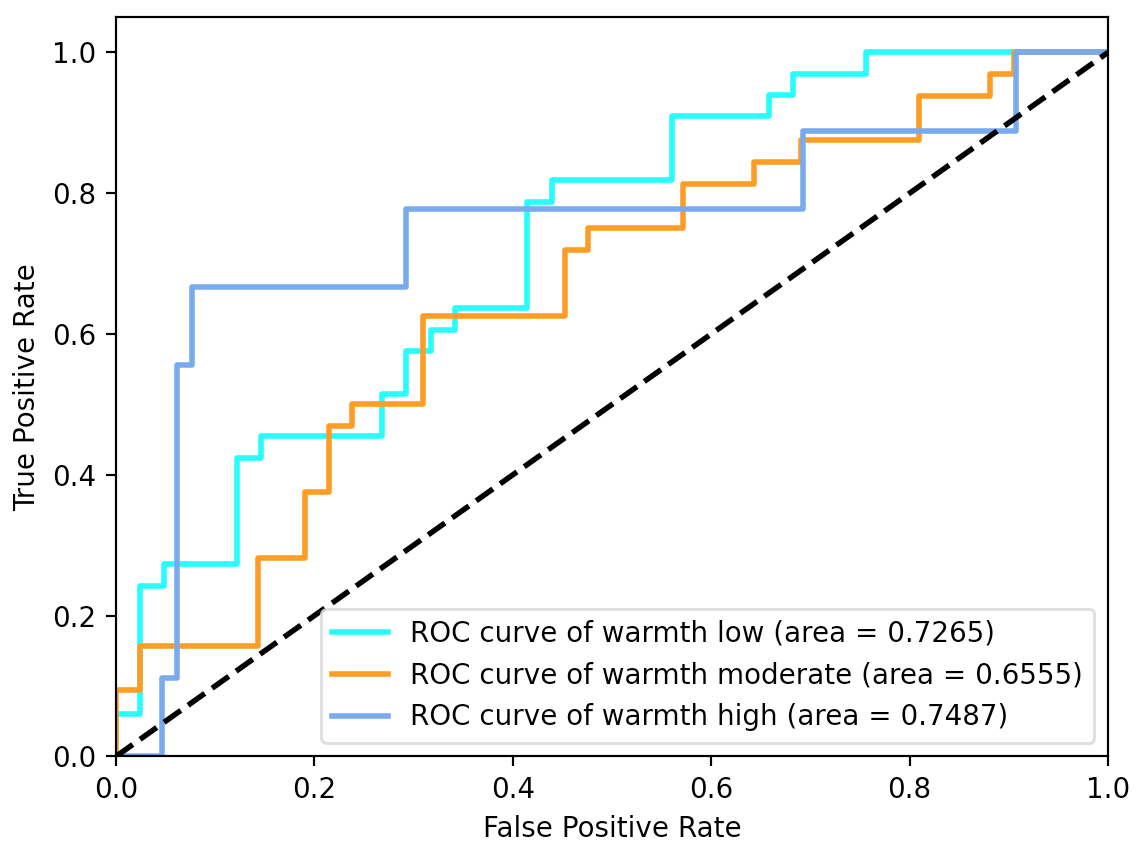} 
 \caption{Receiver operating characteristic curves of the three-way Lin-SVC classification using is10-par acoustic features combined with \tfidfpun text features.}
 \label{fig:roc} 
 
\end{figure}

\begin{figure}[t]
 \centering
 \includegraphics[width=7.25cm]{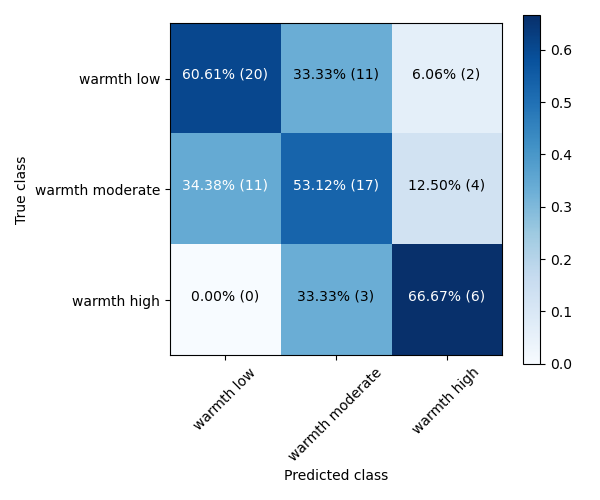} 
 \caption{Confusion matrix of the three-way Lin-SVC classification using is10-par acoustic features combined with \tfidfpun text features.}
 \label{fig:cm} 
 \vspace{-0.4cm}
\end{figure}

Figure~\ref{fig:roc} shows the Receiver Operating Characteristic (ROC) curves for one of the representative runs of the best performing classifier (Lin-SVC) on IS10-CMP combined with \tfidfpun features. Figure~\ref{fig:cm} shows the corresponding Confusion Matrix (CM). The best ROC curve (up to 0.7487) was obtained for \textsl{high} warmth (the minority class). Highest accuracy was also achieved for this class (67\% or 6 out of 9), followed by the \textsl{low} class (the majority class) with accuracy of around 61\% (20/33). The most challenging class to predict was \textsl{moderate} which had the highest degree of confusion (with \textsl{low} warmth) and an accuracy of 53\%.


\section{Conclusions}\label{conclusion}

This study has highlighted some of the significant challenges of working with a dataset of \textsl{real-world} audio. First, the low quality of audio recorded on cassette tapes and low-grade equipment greatly complicated the task of transcription and significantly increased the investment required to make the dataset usable. Second, the cost of transcription and a restricted project budget limited the number of transcripts we could obtain. Manually tagging the utterances in the interviews proved an effective way of increasing the amount of data and enabled us to split interviews into two parts, one for each twin. Third, the small amount of available data imposed limitations on the choice of classification models we could use. It was not feasible to trial state-of-the-art models that require much larger amounts of data, such as deep neural networks. Finally, due to class imbalance, it was necessary to merge the classes of our target variable. Our experiments should, therefore, be seen as a proxy for the task of predicting true warmth scores as coded in the E-Risk study.

Despite these limitations, our results tentatively indicate that combining acoustic and text features is optimal when trying to predict the levels of caregiver warmth expressed in Five Minute Speech Samples. This promising result suggests that machine learning classifiers may eventually be an adequate substitute for the process of manual coding of warmth, and EE more generally, by human raters.

In future work, we intend to prioritise the expansion of the dataset with additional transcriptions. A larger dataset would open up the possibility of using more sophisticated classification models. Ultimately, however, we aim to develop an approach based on automatic speech recognition in order to alleviate the burden of manual transcription.

We conclude by making recommendations for researchers faced with the challenges of working with \textit{real-world} speech samples. Investing time in the manual preparation of data, such as tagging or other annotations, can help mitigate the effects of limited and low quality data. We suggest adapting experiments to the limitations of the data by using a variety of established and recent feature extraction and machine learning methods. Finally, in initial experiments consider using a combination of a reduced number of classes and conventional machine learning approaches, as this should help keep model complexity lower when only smaller amounts of data are available.





\section{Acknowledgements}
We are grateful to the E-Risk study families and twins for their participation. Our thanks to Professors Terrie Moffitt and Avshalom Caspi, the founders of the E-Risk study, and to the E-Risk team for their dedication, hard work, and insights. The E-Risk Study is funded by the Medical Research Council (UK MRC) [G1002190]. This project was funded by the Psychiatry Research Trust [39C]. JD received support from a National Institute of Health Research (NIHR) Clinician Scientist Fellowship [CS-2018-18-ST2-014] and Psychiatry Research Trust Peggy Pollak Research Fellowship in Developmental Psychiatry. HLF is part supported by the Economic and Social Research Council (ESRC) Centre for Society and Mental Health at King's College London [ES/S012567/1]. The views expressed are those of the authors and not necessarily those of the ESRC, NIHR, the Department of Health and Social Care, or King’s College London.

\end{document}